\documentclass[9pt,twocolumn,twoside]{pnas-new-no-watermark}
\templatetype{pnasresearcharticle-v1}
\usepackage{mathtools}
\usepackage{float}
\usepackage{multirow}
\usepackage{ucs,amssymb,amsmath,mathtools}
\usepackage[utf8]{inputenc}
\usepackage{fontenc}
\usepackage{graphicx}
\usepackage[pdftex,colorlinks]{hyperref,ragged2e}
\usepackage[
singlelinecheck=false % <-- important
]{caption}

\DeclareMathOperator*{\argmin}{argmin}

\usepackage{setspace}
\usepackage[normalem]{ulem}

\newcommand{\Nb}{ \nobreak}

\newcommand{\Pv}{P\nobreakdash-value}

\newcommand{\Pvs}{P\nobreakdash-values}
\newcommand{\OR}{\text{OR}}
\newcommand{\RR}{\text{RR}}

\title{Maximum value of the standardized log of odds ratio and celestial mechanics}

\author[a]{Olga A. Vsevolozhskaya}
\author[b]{Gabriel Ruiz}
\author[c,1]{Dmitri V. Zaykin}

\affil[a]{Department of Biostatistics, University of Kentucky, Lexington, USA}
\affil[c]{National Institute of Environmental Health Sciences, National Institutes of Health, USA}
\affil[b]{University of California, Los Angeles, USA}

\leadauthor{Vsevolozhskaya} 

\significancestatement{Logarithm of the odds ratio, $\mu$, divided by its asymptotic standard deviation, $\sigma$, gives a standardized quantity, $\gamma$=$\mu$/$\sigma$. It is connected to statistical power, because $\gamma$ times the square root of sample size is the noncentrality parameter of the asymptotically normal statistic. Here we show that the maximum value of $\gamma$ is the Laplace Limit Constant, 0.6627... that governs the convergence of the classic power series solution to the Kepler Equation, a widely recognized planetary motion equation.}

\authordeclaration{The authors declare no conflict of interest here.}
\correspondingauthor{\textsuperscript{1} Corresponding author (e-mail: dmitri.zaykin@nih.gov)}

\begin{abstract} %250 words max
The odds ratio (OR) is a widely used measure of the effect size in observational research. ORs reflect statistical association between a binary outcome, such as the presence of a health condition, and a binary predictor, such as an exposure to a pollutant. Statistical significance and interval estimates are often computed for the logarithm of OR, ln(OR), and depend on the asymptotic standard error of ln(OR). For a sample of size $N$, the standard error can be written as a ratio $\hat{\sigma}/\sqrt{N}$, where $\sigma$ is the population standard deviation of ln(OR). The ratio of ln(OR) over $\sigma$ is a standardized effect size. Unlike correlation, that is another familiar standardized statistic, the standardized ln(OR) cannot reach values of minus one or one. We find that its maximum possible value is given by the Laplace Limit Constant, (LLC=0.6627...), that appears as a condition in solutions to Kepler equation -- one of the central equations in celestial mechanics. The range of the standardized ln(OR) is bounded by minus LLC to LLC, reaching its maximum for ln(OR)$\approx$4.7987. This range has implications for analysis of epidemiological associations, affecting the behavior of the reasonable prior distribution for the standardized ln(OR).
\end{abstract}
\begin{document}

% \verticaladjustment{-2pt}
\maketitle

\thispagestyle{firststyle}

\ifthenelse{\boolean{shortarticle}}{\ifthenelse{\boolean{singlecolumn}}{\abscontentformatted}{\abscontent}}{}

% \section*{Introduction}

\dropcap{W}hen both exposure and disease outcome are binary variables, epidemiological data can be conveniently summarized by a 2$\times$2 table:

\begin{table}[ht!]
\centering
%\caption{2 $\times$ 2 table}
\begin{tabular}{ccc}
\hline
&\multicolumn{2}{l}{Exposure}\\
%\vspace{1mm}
\cline{2-3} \rule{0pt}{3ex}
Disease status & $E$ & $\bar{E}$ \\
\hline \rule{0pt}{3ex}
$D$ & $n_{11}=n_D\hat{p}$ & $n_{12}(1-\hat{p})$\\
$\bar{D}$ & $n_{21}=n_{\bar{D}}\hat{q}$ & $n_{22}(1-\hat{q})$\\
\hline
\end{tabular}
\label{tab1}
\end{table}
\noindent where $n_{11}+n_{12}$ is the number of cases, $n_D$; $n_{21}+n_{22}$ is the number of controls, $n_{\bar{D}}$; and the number of exposed subjects is $n_{11}+n_{21}$. When sampling is random with respect to exposure $E$,
sample proportions $\hat{p}=n_{11}/n_D$ and $\hat{q}=n_{21}/n_{\bar{D}}$ estimate population probabilities of exposure among cases and among controls, respectively ($p = \Pr(E|D)$ and $q = \Pr(E|\bar{D})$). Then, in epidemiological studies, the effect of exposure on outcome is often measured by odds ratio, OR, which is defined as:

\begin{eqnarray*}
   \OR &=& \frac{p/(1-p)}{q/(1-q)} \\
 &=& \frac{\Pr(D \mid E)/(1-\Pr(D \mid E)) }{\Pr(D \mid \bar{E})/(1-\Pr(D \mid \bar{E})}.
\end{eqnarray*}
Relative risk, $\RR\Nb=\Nb\Pr(D|E)\Nb/\Nb\Pr(D|\bar{E})$ cannot be directly estimated from table counts when sample proportions of cases and controls are fixed by design, but $\OR$ estimate, $\widehat{\OR}\Nb=\Nb \frac{\hat{p}/(1-\hat{p})}{\hat{q}/(1-\hat{q})}$, is unaffected by the study design. 

Let $\mu$ denote the effect size measured by log odds ratio. Given the estimated log odds ratio, $\hat{\mu} = \ln(\widehat{\text{OR}})$, a commonly used statistic is:
\begin{eqnarray*}
   T &=& \frac{\ln(\widehat{\text{OR}})} {\sqrt{\sum{1/n_{ij}}}} = \frac{\hat{\mu}}{\sqrt{\sum{1/n_{ij}}}},
\end{eqnarray*}
which asymptotically follows the standard normal distribution. 
The sum of four cell counts, $N=\sum n_{ij}$, can be factored into this expression as:
\begin{eqnarray*}
   T &=& \sqrt{N} \, \, \frac{\hat{\mu}} {\hat{\sigma}(\hat{w})} \\
  \hat{\sigma}(\hat{w}) &=& \sqrt{\frac{1}{\hat{w}} \frac{1}{\hat{p}(1-\hat{p})} + \frac{1}{1-\hat{w}} \frac{1}{\hat{q}(1-\hat{q})}},
\end{eqnarray*}
where $\hat{w}$ is the proportion of cases $n_D/N$. % Alternatively, one can express $w$ in terms of the ratio of cases proportion to controls proportion ($\delta = (n_D/N) / (n_{\bar{D}}/N)$) as $w=\delta/(1+\delta)$. 
The corresponding population parameter can be written as:
\begin{eqnarray}  
    \sigma^2(w) &=& \frac{1}{w} \frac{1}{\Pr(E|D)\left[1-\Pr(E|D)\right]}  \label{eq:sd}   \\  \nonumber
  &+& \frac{1}{(1-w)} \frac{1}{\Pr(E|\bar{D})\left[1-\Pr(E|\bar{D})\right]},
\end{eqnarray}
where $w=\Pr(D)$ is disease prevalence. % The ratio $p/q$ has the relation to population prevalence of disease, $\Pr(D)$ and to the risk of disease given exposure, $\Pr(D \mid A)$:
% \begin{eqnarray}
%      \frac{p}{q} &=& \frac{\Pr(A \mid D)}{\Pr(A \mid \bar{D})}
%           = \frac{1-\Pr(D)}{\Pr(D)} \frac{\Pr(D \mid A)}{1-\Pr(D \mid A)}
% \end{eqnarray}
We express variance as a function of $w$ to emphasize that $\sigma(w)$ will vary depending on the study design. Further, solution to $\sigma'(w)=0$, under the constraint $0<w<1$, provides the value of $w$, at which variance is minimized, and thus $\mu / \sigma$ value is maximized. This minimization value can be found as:
\begin{eqnarray}
   w_{m} &=& \argmin_w \sigma(w) = \frac{1}{1 + \frac{\Pr(E|D)}{\Pr(E|\bar{D})} \,\, \sqrt{\text{OR}^{-1}} }. \label{wm}
\end{eqnarray}
Thus, the ratio $\gamma  = \mu / \sigma$ will attain its maximum if $\sigma = \sigma(w_m)$. Alternatively, in terms of the pooled exposure probability,  $v=w\Pr(E|D)+(1-w)\Pr(E|\bar{D})$, the value at which variance is minimized and $\mu / \sigma$ is maximized can be expressed as a function of RR and OR as:
\begin{eqnarray}
  v_m &=& \argmin_v \sigma(v) = \frac{1}{1 + \text{RR} \,\, \sqrt{\text{OR}^{-1}} }. \label{vm}
\end{eqnarray}
The variance of the prior distribution for $\mu / \sigma(v)$ will thus reach its minimum at $\sigma = \sigma(v_m)$.
Re-expressing $\sigma$ in Eq. (\ref{eq:sd}) as a function of $v$, we get:
\begin{eqnarray}
  \sigma^2(v) &=& \frac{1}{v} \frac{1}{\Pr(D|E)\left[1-\Pr(D|E)\right]} \label{sigma.vm} \\  \nonumber
  &+& \frac{1}{1-v} 
      \frac{1}{\Pr(D|\bar{E})\left[1-\Pr(D|\bar{E})\right]}. 
\end{eqnarray}
To obtain maximum possible standardized $\ln(\OR)$, we can substitute $v_m$ and $\Pr(D|\bar{E})=1/(1-\OR\left[1-1/\Pr(D|E)\right])$ into Eq. (\ref{sigma.vm}), and minimize the resulting equation with respect to exposure risk, $\Pr(D|E)$, and with respect to $v$, which results in:
\begin{equation}
  \Pr(D|E)= 1 - \frac{1}{1 + \sqrt{OR}}, \label{eq5}
\end{equation}
\begin{equation}
\Pr(D|\bar{E}) = \frac{1}{1 + \sqrt{\text{OR}}} = 1 - \Pr(D|E), \label{eq6}
\end{equation}
and
\begin{equation}
  v = 1/2.  \label{eq7}
\end{equation}
Next, by substituting Eqs. (\ref{eq5}-\ref{eq7}) into Eq. (\ref{sigma.vm}) we get the denominator of the maximum standardized effect size. Therefore:
\begin{eqnarray}
   \gamma_{\max} = \frac{ \ln(\OR) }{ 2 \sqrt{2 + (1+\OR) / \sqrt{\OR}} }. \label{gamma.max}
\end{eqnarray}
% This equation which we recently stated but gave no details\cite{vrz2017bayesian} 
The above equation depends only on odds ratio but is not monotone in it, and reaches its maximum for ln(OR) value about 4.7987. Perhaps counterintuitively, but as ln(OR) exceeds that value, the corresponding standardize statistic, ln(OR)/$\sigma$, starts to decrease.

\begin{figure}[h!]
\centering
\includegraphics[width=0.8\linewidth]{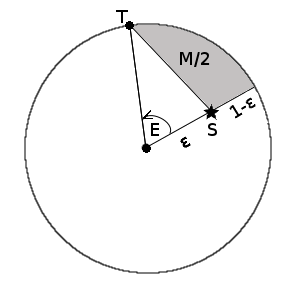} 
\caption{\textbf{The Kepler equation: geometric interpretation}\\ Given the knowledge of the area $M$ and the distance to the origin, $\varepsilon$, solve for the angle $E$ in $M = E - \varepsilon \sin(E)$.\cite{colwell1993solving}
}
\label{fig1}
\end{figure}

It turns out that there is a peculiar connection between the expression for $\gamma_{\text{max}}$ and the famous orbital mechanics equation: the Kepler equation, $M = E - \varepsilon \sin(E)$. A geometric interpretation of the Kepler equation is illustrated by Figure \ref{fig1}. Suppose that we are inside a circular orbit rescaled to be the unit circle. Our position S is denoted by ``$\large{\star}$''. The shortest path to the orbit has the length $1-\varepsilon$. A celestial body traveles the orbit from that point to T. Given the area $M/2$ and distance $1-\varepsilon$, we want to determine the angle $E$. These three values are related to one another by Kepler's equation. Planetary orbits are elliptical, so the actual orbit is along an ellipse inside of the unit circle% , therefore, T can be thought of as a projection of the planet's actual location onto the circle
. Still, the calculation of the {\em eccentric anomaly} $E$ is a crucial step in determining planet's coordinates along its elliptical orbit at various times.
The Kepler equation (KE) is transcendental, i.e., with no algebraic solution in terms of $M$ and $\varepsilon$, and it has been studied extensively since it is central to celestial mechanics. Colwell writes ``The sole subject of our work is Kepler's Equation'' in the book suitably named ``Solving Kepler's equation over three centuries''  and notes that ``in virtually every decade from 1650 to the present'' there have been papers devoted to that equation.\cite{colwell1993solving} A solution to KE can be written as an infinite series in powers of $\varepsilon$, which is convergent only if $\varepsilon$ is smaller than the ``Laplace Limit Constant'', LLC. To relate LLC to the bounds for standardized $\ln(\OR)$, let $x = \ln(\OR)$, $x>0$. In terms of $x$, the standardized statistic is given by:
\begin{eqnarray}
   \gamma = \kappa(x) = \frac{x}{2 \sqrt{2 + \frac{1 + \exp{(x)}}{\exp{(x/2)}}}}.
\end{eqnarray}
Using basic trigonometric identities:
\begin{eqnarray*}
   && \frac{1 + \exp{(x)}}{\exp{(x/2)}} = 2 \cosh (x/2), \\
   && 2 + \frac{1 + \exp{(x)}}{\exp{(x/2)}} = 4 ( \cosh (x/4) )^{2}, \quad \text{and} \\
   && \sqrt{2 + \frac{1 + \exp{(x)}}{\exp{(x/2)}}} = 2 \cosh (x/4), 
\end{eqnarray*}
we can express $\kappa(x)$ and its derivative in terms of hyperbolic functions as:
\begin{eqnarray}
   \kappa(x) &=& \frac{x/4}{\cosh(x/4)} = (x/4) \, \text{sech}(x/4) \label{eq.kappa} \\ 
   \kappa'(x) &=& \frac{4 - x \tanh(x/4) }{16 \cosh(x/4) }.
\end{eqnarray}
To maximize the standardized ln(OR), we set $\kappa'(x)$=$0$, which is equivalent to solving $(x/4)\tanh(x/4)=1$. The solution is four times the solution to $x\tanh(x)=1$ equation, which is 1.19967864... This implies maximum ln(OR) $= 4 \times 1.19967864...=4.7987...$ and by substituting this value into Eq. (\ref{gamma.max}) we obtain $\gamma_{\text{max}} = 0.6627...$, the LLC. The solution to KE % can be expressed as the power series in $\varepsilon$, provided $|\varepsilon \sin(E)| < |E-M|$ and that
involves the condition equivalent to Eq. (\ref{eq.kappa}). Namely, the solution can be expressed as the power series in $\varepsilon$, provided $|\varepsilon \sin(E)| < |E-M|$ and that $\varepsilon < x / \cosh(x), x = |E-M|$, which is the LLC.\cite{plummer1918introductory}

Although it appears that the LLC bound is a function of odds ratio alone, this bound can only be attained at the specific values of population parameters (or the respective sample values). Namely, (i) $v_m=w_m=1/2$ from Eq. (\ref{eq7}), which implies $\RR^2= \left(\frac{\Pr(E|D)}{\Pr(E|\bar{D})}\right)^2 = \OR$; (ii) $\Pr(D|E)=1-\Pr(D|\bar{E})$ from Eqs. (\ref{eq5} - \ref{eq6}); and (iii) ln(OR) $=  4.7987...$ Next, by solving
\begin{eqnarray*}
   \OR &=& \frac{\Pr(D|E)/(1-\Pr(D|E))}{\Pr(D|\bar{E})/(1-\Pr(D|\bar{E}))} \\
   &=& \exp(4.7987\dots) = 121.354\dots
\end{eqnarray*}
for $\Pr(D|E)$, we obtain:
\begin{eqnarray}
   \Pr(D|E) = \frac{1}{2\,z} + \frac{1}{2},
\end{eqnarray}
where $z$ is the solution  of $ z  \tanh(z)=1$, i.e., $z =1.19967864\dots$ and $\Pr(D|E) = \Pr(E|D) = 0.9167782798\dots$
Similarly, Pearson correlation between two binary variables ranges between -1 and 1, but this range is not free of parameters: these boundary values are possible only in the case when the population (or sample) frequencies of two binary variables are equal to each other. Moreover, these bounds are asymmetric, depending on the sign of the correlation.\cite{weir1979inferences} 

The range of the standardized statistic, -LLC to LLC, has implications for statistical analysis. For example, several recent publications on \Pv{} replicability posed the following question: given a small initial \Pv{}, what is a likely spread of \Pvs{} in subsequent replication studies?\cite{halsey2015fickle,lai2012subjective,lazzeroni2014p,lazzeroni2016solutions,vrz2017bayesian} \Pvs{} for $\ln(\OR)$ are explicit functions of $\gamma$ because they are defined as $P = \Pr(Z  > z_{\alpha})$, where $Z = \sqrt{N} \gamma$ is asymptotically normal. The two-sided \Pv{} can be similarly defined in terms of chi-square distributed $Z^2$. The prior distribution for the standardized effect size occurs naturally and needs to be specified in order to give probabilistic bounds for the spread of future replication \Pvs{}. In applications where prior distribution for the effect size is modeled in terms of ln(OR), % the function $\kappa(x)$ and its
our results allow one to specify a reasonable prior range for the standardized value.

 It has been suggested that summary association statistics can be converted to approximate posterior (Bayesian) summaries about parameters of interest. For example, one-sided \Pv{}, $P$ for testing significance of $\ln(\OR)$ can be transformed to the normal test statistic, $Z = \Phi^{-1}(1-P)$. This statistic is $Z=\sqrt{N}\ln(\widehat{\OR})/\hat{\sigma}$. An approximate Bayesian false discovery probability can be computed based only on the summary statistics $\ln(\widehat{\OR})$ and $\hat{\sigma}$, the value $N$, and an assumed variance parameter for the zero-mean prior normal distribution for $\ln(\OR)$.\cite{wakefield2007bayesian,wakefield2009bayes} For any given value of OR,  $\mu=\ln(\OR)$, is fixed, but $\sigma$ can vary as a function of $w$. The normal prior distribution for $\mu$ can be characterized simply by $\Pr(\OR > x) = \beta$. Considering the standardized effect, we can write $\beta = \Pr(\OR > x) = \Pr(\mu / \sigma > \ln(x) / \sigma)$.
Denote the normal cumulative distribution function with the mean $a$ and variance $b$, evaluated at $x$ by $\Phi(x|a,b)$, and its inverse by $\Phi^{-1}(x|a,b)$. Then,
$\ln(x) /\sigma = \Phi^{-1}(1-\beta \mid 0,\sigma_0) = \sqrt{\sigma_0} \Phi^{-1}(1-\beta \mid 0,1)$.
From this, we can obtain the flattest possible prior distribution for $\mu/\sigma$ as the zero-mean normal with variance
\begin{eqnarray}
   \sigma_0 = \left( \frac{\ln(x)/\sigma_m}{\Phi^{-1}(1-\beta \mid 0,1)} \right).
\end{eqnarray}
For $\sigma_0$ to be as large as possible, $\sigma_m$ should be equal to $\ln(x)/\gamma_{\max}$ (from Eq. \ref{gamma.max}). Alternatively, either the value $\sigma(w_m)$ or $\sigma(v_m)$ can be specified with some additional assumptions. For example, for $\sigma(w_m)$, $\Pr(D|E)$ needs to be specified. Then,
\begin{eqnarray}
   \Pr(D \mid \bar{E}) &=& \frac{1}{1-\OR \left(1 - \Pr(D \mid E)^{-1} \right)}, \\
   \RR &=& \Pr(D \mid E) /  \Pr(D \mid \bar{E}),
\end{eqnarray}
and $\sigma(w_m)$ is obtained using the value $w_m$ from Eq. (\ref{wm}). \\

%\begin{eqnarray}
%   \ln(\OR) = 2 \arctanh\left(\frac{\OR-1}{\OR+1}\right)
%\end{eqnarray}

\section*{Acknowledgements}
This research was supported in part by the Intramural Research Program of the NIH, National Institute of Environmental Health Sciences.

\bibliography{Kepler.bib}

\end{document}